\documentstyle[eqsecnum,preprint,aps]{revtex}
\tightenlines

\def\beq{\begin{equation}}
\def\eeq{\end{equation}}
\def\bea{\begin{eqnarray}}
\def\eea{\end{eqnarray}}
\def\nnu{\nonumber}
\def\tst{\textstyle}

\def\al{\alpha}
\def\be{\beta}
\def\gam{\gamma}

\def\eps{\epsilon}
\def\tta{\theta}
\def\kap{\kappa}
\def\om{\omega}

\def\Dta{\Delta}
\def\Sig{\Sigma}

\def\sech{{\rm sech\,}}
\def\ptl{\partial}
\def\hf{{1\over2}}
\def\tshf{\tst\hf}

\def\lp{\left(}
\def\rp{\right)}

\def\blp{{\bigl(}}
\def\brp{{\bigr)}}

\def\adag{a^{\dagger}}
\def\ham{{\cal H}}
\def\ket#1{|#1\rangle}

\def\tran#1#2{\langle#1|#2\rangle}

\def\bnk{b_n^{(k)}}
\def\bS{{\bf S}}
\def\cmk{C_m^{(k)}}
\def\D0{\Delta_0}
\def\Dk{\Delta_k}
\def\Da{\Delta_{\rm asymp}}
\def\Di{\Delta_{\rm int}}
\def\e0k{E_0^{(k)}}
\def\hos{\ham_{\rm osc}}
\def\kc{\textstyle{k_1 + \hf k_2}}
\def\sjj{\sum_{j \ge j_0}}
\def\sph{\tst{S + \hf}}
\def\smm{\tst{S + \hf -m}}
\def\spm{\tst{S + \hf +m}}
\def\sti{S \to \infty}
\def\tmid{t_{\rm mid}}
\def\Suto{S\"ut\H o}

\begin{document}
\draft

\title{Application of the discrete WKB method to spin tunneling}

\author{Anupam Garg}
\address{Department of Physics and Astronomy, Northwestern University,
Evanston, Illinois 60208}

\date{today}

\maketitle

\begin{abstract}
A discrete version of the WKB method is developed and applied to
calculate the tunnel splittings between classically degenerate states
of spin Hamiltonians. The results for particular model problems
are in complete accord with those previously found using instanton
methods. The discrete WKB method is more elementary and also yields
wavefunctions.
\end{abstract}
\pacs{03.65.Sq, 33.20.-t, 75.45.+j}

\widetext

\section{Introduction}
\label{Intro}

A large variety of problems in physics can be modeled in terms of a spin
of large magnitude which is then described by a Hamiltonian which is a
polynomial in the spin components. It is natural in many of these cases to
seek a semiclassical description. This is especially so when the system
possesses a number of degenerate states in the classical limit, and one
wishes to see how this degeneracy is lifted by quantum tunneling.
Examples include the Lipkin-Meshov-Glick
model in nuclear physics\cite{lmg,rs,kl}, the
rotational spectrum of polyatomic molecules~\cite{hp}, and the
Hamamoto-Mottleson model~\cite{hm} for the $\Dta I = 4$ staggering of
the rotational spectra of certain superdeformed nuclei.

Perhaps the simplest example of such a problem is provided by the Hamiltonian
\beq
\ham  =  -k_1 S_z^2 + k_2 S_x^2, \label{ham} 
\eeq
where $k_1 > 0$, $k_2 > 0$, and $S_{\al}$ ($\al = x$, $y$, $z$) are
components of the spin operator $\bS$ obeying the usual commutation rules
\beq
[S_{\al}, S_{\be}] = i \eps_{\al\be\gam}S_{\gam}.
\label{cr}
\eeq
Further, the spin has a magnitude $S$, i.e., the operator $\bS\cdot\bS$ has
eigenvalue $S(S+1)$. As $S \to \infty$, the system is more and more classical,
and in the limit it has two degenerate ground states corresponding to
$S_z/S = \pm 1$. For large but finite $S$, the quantal analogs of these
states will be admixed by a tunneling matrix element $\D0/2$. The two
lowest energy eigenstates will then be split by an amount $\D0$. The goal
of a tunneling calculation is to find an expression for $\D0$ that is
asymptotically valid as $S \to \infty$. In analogy with the result for
particles with a position like coordinate in a potential well, we expect this
expression to take the form
\beq
\D0 = e^{-(B_0 + B_1 + B_2 + \cdots)},
\label{asy}
\eeq
where
\beq
B_0 \gg B_1 \gg B_2 \gg \cdots
\label{Bseq}
\eeq
in the sense that $B_1/B_0$, $B_2/B_1, \ldots \to 0$ as $\sti$. By looking at
the commutation relations for the scaled operators $S_{\al}/S$, we can see
that $1/S$ plays the same role as $\hbar$ in conventional WKB theory, and we
thus expect $B_0 = O(S)$. As is often done for a particle in a well, it is
convenient to call $e^{-B_0}$ the Gamow factor, or the WKB exponent, and
the remaining terms $e^{-(B_1 + B_2 + \cdots)}$ the prefactor.

In many physical problems, it is only sensible or possible to calculate
$\D0$ to ``exponential accuracy", i.e., to find only the leading term $B_0$ in
Eq.~(\ref{asy}). As a general point of formalism, however, it is clearly
desirable to have a proper asymptotic approximation $\Dta'_0$ to $\D0$ itself and
not just to $\ln\D0$. In other words, we demand that
$\Dta'_0/\D0 \to 1$ as $S \to \infty$. For this one must know
all $B_n$ that are of $O(S^0)$ or greater.
The first correct calculation to such accuracy that we know of was done by
Enz and Schilling~\cite{es1,es2}, who mapped the spin onto a canonically conjugate
variable pair via a Villain transformation, and then applied an instanton
method~\cite{sc} to the resulting problem.
Great care must be taken in this approach
with operator ordering problems, leading to a rather complex calculation.
A more natural approach is one based on Klauder's spin-coherent-state path
integral~\cite{Kl}. A naive application of instanton methods for the
tunnel splitting to this path
integral \cite{cg}, however, only gives the leading $S$ dependence
correctly. In particular, terms of $O(S^0)$ in $\ln\D0$ are
incorrectly obtained~\cite{km,gk}.  A correct
calculation within this formalism has only recently been done by Belinicher,
Providencia, and Providencia \cite{bpp}, who show that it is necessary to consider
non-differentiable paths by a careful examination of the discrete time version
of the path integral. Once again, the simplicity of the instanton approach is
lost.

It is therefore desirable to have a means of calculating the tunnel splitting using
only elementary methods of analysis. Such an approach is provided by the discrete
WKB method. (An alternative approach, used in Ref. \cite{SWv}, also meets the
criterion of being elementary.) In the present context, the method
consists of writing the
Schr\"odinger equation as a recursion relation for the expansion coefficients
of an energy eigenstate in the $S_z$ basis, and solving this recursion relation
under the assumption that the coefficients in the recursion relation vary slowly
with $m$, the $S_z$ eigenvalue. By and large, the method is a
straightforward generalization of the continuous WKB method~\cite{ll},
and it may also be
viewed as equivalent to the use of semiclassical dynamics for Bloch electrons.

The earliest appearance of this method in a physics context of which we are aware
is by Schulten and Gordon \cite{sg}, who use it to find semiclassical
approximations for the $3j$ (Wigner) and $6j$ (Racah) symbols. 
A cogent review of the method in the context of atomic physics has been given by
Braun \cite{pab}, who also cites its application in a variety of other physics
settings, as well as earlier discussions in Russian mathematics texts. 
It has also been previously applied to spin tunneling problems by van Hemmen
and S\"ut\H o~\cite{vs1,vs2}. These authors have not, however, utilized the method
to its full potential. Specifically, they calculate
$\D0$ for a class of Hamiltonians including
\beq
\ham  =  -\gam S_z^2 - \al S_x,    \label{ham2}
\eeq
but only the leading $S$ dependence
is correctly found. One of the goals of this paper is to show that the prefactor
can also be found correctly to order $S^0$ with little extra labor. (See Sec. V.)

We will illustrate the discrete WKB method by applying it to
Eqs.~(\ref{ham}) and (\ref{ham2}). We will find
the splittings $\Dk$, $k=0, 1, 2, \ldots$ for all low lying pairs of
energy levels for Eq.~(\ref{ham}), and the lowest splitting $\D0$ for
Eq.~(\ref{ham2}). Our answers for $\D0$ agree completely with those of Enz and
Schilling \cite{es2}, or Belinicher, Providencia, and Providencia \cite{bpp},
and we believe they are obtained with much less effort. While the discrete
WKB method does not have the nice geometrical structure of the spin-coherent-state
path integral approach, it provides compensation in that it yields wavefunctions
in the course of the calculation, which may be of further use; for evaluating
matrix elements of perturbations, for example \cite{ag1,ag2}. Other problems
commonly solved by the continuous WKB method, such as escape from a metastable well,
are also amenable to the method, but we do not discuss these here.

The plan of the paper is as follows. In Sec.~II we briefly review the discrete WKB
method. Our discussion is heuristic and physically motivated. See Braun \cite{pab}
for one that is more formal. In Sec.~III we derive a formula [see Eq.~(\ref{edif})]
for the tunnel splitting for a symmetric double well problem analogous to that for
the continuous case~\cite{ll2,vHW}.  We apply the method to the Hamiltonian
(\ref{ham}) in Sec.~IV, and to (\ref{ham2}) in Sec.~V. We compare our answers for
the splittings with numerical results, and in the case of Eq.~(\ref{ham2}) to
answers given by van Hemmen and and S\"ut\H o~\cite{vs2}, and Scharf, Wreszinski, and
van Hemmen~\cite{SWv}.

\section{The discrete WKB method}
\label{dwkb}

To introduce the discrete WKB method, let us consider the Hamiltonian
(\ref{ham}), and focus on an energy eigenfunction $\ket\psi$. Denoting
$S_z$ eigenfunctions $\ket m$ with $S_z\ket m = m \ket m$ as usual, and writing
$C_m = \tran m\psi$, it is easy to see that Schr\"odinger's equation
takes the form of a three-term recursion relation connecting $C_{m-2}$,
$C_m$, and $C_{m+2}$. The step $\Dta m =2$ in the index is inconvenient,
so it is preferable to change the notation somewhat, and relabel the states
which are connected to each other by an index $j$, with 
$j=1, 2, \ldots, N$, where $N$ equals $S$ or $S+1$.
The Schr\"odinger equation then takes the form
\beq
w_j C_j + t_{j,j+1} C_{j+1} + t_{j,j-1}C_{j-1} = E C_j,
\label{ttr}
\eeq
where $w_j = \ham_{jj}$ and $t_{j,j\pm 1} = \ham_{j,j\pm 1}$, with
$t_{N,N+1}$ and $t_{N+1,N}$ being understood as zero.
We may also assume that $t_{j,j+1} \ne 0$ for any $j$, since otherwise the
matrix for $\ham$ is block diagonal, and Eq.~(\ref{ttr}) may be regarded as
applying to one block. We can also choose the phases of the basis states
so that all $t_{j,j\pm 1}$ are real. Hermiticity then yields
$t_{j+1,j} = t^*_{j,j+1} = t_{j,j+1}$.

The above problem is equivalent to a tight-binding model for an electron on a
one-dimensional lattice with on-site energies $w_j$, and hopping matrix
elements $t_{j,j\pm 1}$, and indeed our notation is chosen to reflect this
analogy. If $w_j$ and $t_{j,j+1}$ vary slowly enough with $j$ (how slowly
will be seen below) we may use the approximation of semiclassical electron
dynamics. To do this we first note that for the uniform case where $w_j = w$
and $t_{j,j+1} = t$ for all $j$, the eigenstates are Bloch states with
wavefunction $e^{iqj}$ and energy $E(q) = w + 2 t\cos q$. For the nonuniform
case, we first extend $j$ to be a continuous variable, and define 
$w(j)$ and $t(j)$ to be extensions of $w_j$ and $t_{j,j\pm 1}$ via the
relations
\bea
w(j) & = & w_j, \label{wj} \\
t(j) & = & (t_{j,j+1} + t_{j,j-1})/2, \label{tj}
\eea
which are required to hold whenever $j$ is an integer.
We further demand that $w(j)$ and $t(j)$
be smooth enough that if $j/N$ is regarded as a quantity of order $N^0$, then
\beq
{dw \over dj} = O\left( {w(j) \over N } \rp,
\label{slow}
\eeq
and likewise for $t(j)$. This equation is the formal statement of `slowly varying'
coefficients. Whether or not it can be satisfied depends, of course, on the
problem under consideration, and it is this condition which determines whether or
not the problem is amenable to solution by the discrete WKB method. It is
easy to see that it holds for Eq.~(\ref{ham}) whenever $S \gg 1$.

We next define a local wavevector $q(j)$ by the relation
\beq
\cos q(j) = \blp E - w(j) \brp \big/ 2 t(j).
\label{qj}
\eeq
Further, since the electron velocity is given by $\ptl E/\ptl q$ in semiclassical
dynamics, we define
\beq
v(j) = - 2 t(j) \sin q(j).
\label{vj}
\eeq
Then by a straightforward repeat of the arguments employed in the continuous
WKB method (see \cite{ll}, e.g.), it is a simple matter to show that two
linearly independent solutions to Eq.~(\ref{ham}) are given by
\beq
C_j \sim {1 \over \sqrt{v(j)}} 
              \exp\left( \pm i \int^j q(j') dj' \rp.
\label{Cj}
\eeq
A general solution is obtained by taking linear combinations of these
two solutions.

A formal proof of Eq.~(\ref{Cj}) is given by Braun~\cite{pab}, along with
connection formulas, Bohr-Sommerfeld quantization rules, etc. We will therefore
limit ourselves to a few key comments and comparisons with the continuous WKB
method.

The first point is that Eq.~(\ref{Cj}) represents a development of $\log C_j$
in powers of $N^{-1}$, in which only the leading two terms are kept.
To this accuracy, we need only
keep the first two terms of $w(j)$ and $t(j)$ in an expansion in powers of
$1/N$. Alternatively, we can use any expression for these functions which will
reproduce the first two terms correctly. In particular, Eqs.~(\ref{wj}) and\
(\ref{tj}) need only hold to this order. In considering this point, it should be
remembered that $j/N$ is regarded as being of $O(1)$.

The second point is that the factor $1/\sqrt v$ in Eq.~(\ref{Cj}) has the effect
of normalizing each of the two solutions so that the probability
flux is conserved and has a value of unity. This follows from noting
that $|C_j|^2 \sim 1/v$, so that the particle spends a time inversely
proportional to its velocity in any given coordinate segment $dj$. Since this
is exactly what we expect for a classical particle, the quantum mechanical
wavefunction is correctly adjusted to give a conserved probability flux.

The third point is that for a given energy $E$, the classically accessible
region of motion is defined by the inequalities
\bea
&{}&U^-(j) \le E \le U^+(j), \label{class} \\
\noalign{\hbox{where}}
&{}&U^{\pm} = w(j) \pm 2|t(j)|.  \label{Upm}
\eea
This follows from Eqs.~(\ref{qj}) and (\ref{vj}) because whenever the condition
(\ref{class}) is violated, $q(j)$ and $v(j)$ acquire imaginary parts. The
expression (\ref{Cj}) [with $\sqrt{v(j)}$ replaced by $\sqrt{|v(j)|}$ as usual]
then describes an exponentially decaying or growing wavefunction as opposed
to an oscillatory one. It may also be noted that for uniform $w_j$ and
$t_{j,j\pm 1}$, the quantities $U^{\pm}$ would be the limits of the allowable
range of Bloch state energies. More generally they may be regarded as local,
$j$-dependent band edges.

Finally, let us note the conditions for the validity of Eq.~(\ref{Cj}). The
physical requirement is that the wavelength of the particle should not change
by very much over dimensions of the wavelength itself, which is equivalent
to demanding that
\beq
|dq(j)/dj| \ll \sin^2q(j).
\label{slowq}
\eeq
In addition to failing if $w(j)$ and $t(j)$ do not vary slowly (as defined above),
it is easy to see that Eq.~(\ref{slowq}) is violated whenever $q \approx 0$ or
$\pm \pi$. In light of Eqs.~(\ref{class}) and (\ref{Upm}), this is hardly
surprising, since these are turning points of the classical motion. Solutions of
the form (\ref{Cj}) on opposite sides of these points must be related to each
other by the above mentioned connection formulas, as in the usual WKB method.

\section{General formula for tunnel splitting}

In this section we consider the problem where the functions $w_j$ and $t_{j,j+1}$
possess reflection symmetry about some point, and the central region is classically
forbidden for some energies. The band-edge functions $U^{\pm}(j)$ then have the
form shown in Fig. 1. We will derive a formula [see Eq.~(\ref{edif})] for the tunnel
splitting between states formed from symmetrical states localized in the two
potential wells. Our approach is completely analogous to that for a continuous
coordinate \cite{ll2}. Except for the extension to odd and even numbers of
sites, essentially the same derivation also appears in Ref. \cite{vhsnato},
and a formula very close to Eq.~(\ref{edif}) also appears in van Hemmen and
Wreszinski~\cite{vHW}.  Our notation is sufficiently different, and the
argument sufficiently brief, that it is worth giving it in toto.

It is convenient to shift the index $j$ so that it runs over the values
\beq
-{N-1 \over 2}, -{N-3 \over 2},\ldots, {N-1 \over 2}.
\label{Nval}
\eeq
Note that $j$ takes on integer values when $N$ is odd, and half-integer values
when $N$ is even. With this labelling, we have
\beq
     t_{j,k} = t_{-k,-j} = t_{k,j}.
\label{tsym}
\eeq
It is also convenient, for future use, to define $j_0 = 1$ or $1/2$, depending
on whether $N$ is odd or even, respectively. In either case, $j_0$
labels the first site to the right of the mid-point. We also define $J = (N-1)/2$,
and $\tmid = t_{10}$ or $t_{1/2, -1/2}$ for odd or even $N$.

Let $C_j$ be the normalized wavefunction localized in the right well, and which
satisfies the Schr\"odinger equation (\ref{ttr}), neglecting the possibility
of tunneling, with some energy $E_0$. Further, let $a_j$ and $s_j$ be the
exact associated antisymmetric and symmetric wavefunctions, respectively, and
let them have energies $E_1$ and $E_2$, where $(E_1 + E_2)/2 = E_0$. These
functions are given to very good accuracy by
\beq
\begin{array}{rcl}
a_j & = & {\displaystyle{1\over \sqrt 2}} (C_j - C_{-j}), \\
\noalign{\vskip5pt}
s_j & = & {\displaystyle{1\over \sqrt 2}} (C_j + C_{-j}). \\
\end{array}
\label{ajsj}
\eeq
Since the product $C_j C_{-j}$ is everywhere exponentially small, these functions
are properly normalized to unity.

The Schr\"odinger equations obeyed by $C_j$ and $a_j$ are
\bea
&(w_j - E_0) C_j + t_{j,j-1} C_{j-1} + t_{j,j+1} C_{j+1} = 0, \label{Cj0} \\
&(w_j - E_1) a_j + t_{j,j-1} a_{j-1} + t_{j,j+1} a_{j+1} = 0. \label{aj0}
\eea
We multiply Eq.~(\ref{Cj0}) by $a_j$, Eq.~(\ref{aj0}) by $C_j$, subtract the
latter from the former, and sum the result from $j=j_0$ to $J \equiv (N-1)/2$.
This yields
\beq
(E_1 -E_0) \sjj C_j a_j + \Sig_1 - \Sig_2 = 0,
\label{e1e0}
\eeq
where
\bea
\Sig_1 &=& \sjj a_j(t_{j,j-1} C_{j-1} + t_{j,j+1} C_{j+1}), \label{sig1}\\
\Sig_2 &=& \sjj C_j(t_{j,j-1} a_{j-1} + t_{j,j+1} a_{j+1}). \label{sig2}
\eea
By shifting the index of summation in Eq.~(\ref{sig2}) we can see that most of
the terms in the sums $\Sig_1$ and $\Sig_2$ are identical. Recalling that
$t_{J,J+1} =0$, and that $a_0 = 0$ for odd $N$, we obtain
\beq
\Sig_1 - \Sig_2 = \cases{
   a_1 \, \tmid \, C_0 & (odd $N$); \cr
   a_{1/2} \, \tmid \, (C_{1/2} + C_{-1/2}) & (even $N$). \cr}
\label{sigd}
\eeq

To evaluate the sum $\sum_j a_j C_j$ in Eq.~(\ref{e1e0}), we use Eq.~(\ref{ajsj}).
Since $C_{-j}$ is exponentially small everywhere in the right hand well, and since
$C_j$ is normalized,
\beq
\sjj C_j a_j \approx {1 \over \sqrt2} \sjj C_j^2
                      =      {1 \over \sqrt2}.
\label{norm}
\eeq

Substituting Eqs.~(\ref{ajsj}), (\ref{sigd}), and (\ref{norm}) in
Eq.~(\ref{e1e0}), and recalling that $E_1 - E_2 = 2(E_1 - E_0)$, we obtain~\cite{ll2}
\beq
E_1 - E_2 = \cases{
             -2 \, \tmid \, C_0 (C_1 - C_{-1}) & (odd $N$); \cr
             -2 \, \tmid \, (C^2_{1/2} - C^2_{-1/2}) & (even $N$). \cr}
\label{edif}
\eeq

Equation (\ref{edif}) is the sought for formula for the splitting.
Note that unlike the case of a particle with a continuous coordinate degree of
freedom, it is not now necessary for the antisymmetric state to have the higher
energy. In fact, it is not hard to show that
\beq
{{\rm sgn}}\,(E_1 - E_2) = \cases{
                 1 & (odd $N$); \cr
                 -{{\rm sgn}}\,t & (even $N$). \cr}
\label{sgn}
\eeq

\section{Tunnel splittings for model Hamiltonian (1.1)}
We now apply our general formalism to the Hamiltonian (\ref{ham}). Writing a
general eigenstate of $\ham$ as
\beq
\ket\psi = \sum_m C_m \ket m,
\label{psi}
\eeq
where $S_z\ket m = m \ket m$, Schr\"odinger's equation becomes
\beq
(w_m - E)C_m + t_{m,m+2}C_{m+2} + t_{m,m-2}C_{m-2} = 0,
\label{Seq}
\eeq
with
\bea
w_m &=&  \blp\kc\brp \blp S(S+1) - m^2 \brp, \label{wm} \\
t_{m,m+2} &=& {\tst{1 \over 4}} k_2
              \bigl[ \blp S(S+1) - m(m+1) \brp
                     \blp S(S+1) - (m+1)(m+2) \brp \bigr]^{1/2}. \label{tm}
\eea
For convenience, we have added a constant $k_1S(S+1)$ to the Hamiltonian,
and thus to $w_m$.

As noted earlier, Eq.~(\ref{Seq}) is of the same form as Eq.~(\ref{ttr}) with
the minor difference that the index $m$ jumps in steps of 2 rather than 1.
We can take care of this point by changing the integral in
the exponential in Eq.~(\ref{Cj}) to $\int q(m') dm'/2$, leaving the other formulas
unaffected.

One elementary point should be noted at this stage. For half-integer $S$,
Kramers' theorem ensures an exact double degeneracy of all energy eigenstates
of the Hamiltonian (\ref{ham}). This means that $\Dk$ vanishes. In the
spin-coherent-state path
integral approach, this result comes about because the kinetic term in the
action is a Berry phase, and different symmetry related paths acquire phases
which give rise to destructive interference for half-integer
$S$ \cite{ldg,vdh,ag3}. In the discrete WKB method, this result comes about
from an almost trivial separation of the eigenvalue problem into two isomorphic
disjoint subspaces.

There are two steps in the calculation of $\Dk$ via Eq.~(\ref{edif}). First, one
must find the wavefunction $C_m$ in the classically allowed and disallowed
regions separately. In the present case, an expression for the former can be
found exactly and in such a way that it actually holds in part of the
disallowed region. This allows one to match on to the quasiclassical wavefunction
given by Eq.~(\ref{Cj}) without using connection formulas. The
second step consists of substituting the quasiclassical wavefunction into
Eq.~(\ref{edif}).

To execute the first step, let us consider a wavefunction $\cmk$ with energy
$\e0k$ which corresponds to the {\it k\/}th state localized in the well near
$m = -S$, and begin by finding it near this well. The coefficients $w_m$ and
$t_{m,m+2}$ are evaluated by writing $m = -S + n$, and expanding
Eqs.~(\ref{wm}) and (\ref{tm}) in powers of $1/S$, with $n$ regarded as a quantity
of order $S^0$. This yields
\bea
w_{-S+n} &=& \blp\kc\brp (2n+1) S, \label{wn} \\
t_{-S+n,-S+n+2} &=& {\tshf} k_2 S \bigl( (n+1)(n+2) \bigr)^{1/2}. \label{tn}
\eea

Rewriting $C_{-S+n} = b_n$, the recursion relation (\ref{Seq}) reads
\beq
(2k_1 + k_2) S(n+{\tshf}) b_n + {\tshf} k_2 S\sqrt{n(n+1)} b_{n-2}
          + {\tshf} k_2 S \sqrt{(n+1)(n+2)} b_{n+2} = E b_n,
\label{eqbn}
\eeq
which is immediately recognizable as arising from the harmonic oscillator Hamiltonian
\beq
\hos = (2k_1 + k_2) S \left(\adag a + {\tshf}\rp +
              {\tshf} k_2 S \left( a^2 + (\adag)^2 \rp,
\label{hos}
\eeq
where $a$ and $\adag$ are the usual annihilation and creation operators obeying
$[a,\adag] =1$. To see this let us denote the {\it k\/}th energy state of $\hos$ by
$\ket{\psi_k}$, and the number eigenstates by $\ket n$ with $\adag a \ket n = n\ket n$.
It is then apparent that $\bnk = \tran n{\psi_k}$ obeys the eigenvalue equation
(\ref{eqbn}).

To find $\bnk$, we transform to
position and momentum operators $x$ and $p$ obeying
$[x,p] =i$ via $a = (x+ip)/\sqrt2$, $\adag = (x-ip)/\sqrt2$. In terms of these,
\beq
\hos = \left( \kc\rp S (x^2 + p^2) + \tshf k_2 S(x^2 - p^2).
\label{hos2}
\eeq
It follows that the {\it k\/}th state $\ket{\psi_k}$ has an energy
\bea
\e0k   &=& (k + {\tshf}) \om_0, \label{e0k} \\
\om_0  &=&  2S \bigl( k_1(k_1 + k_2) \bigr)^{1/2}. \label{om0}
\eea
The wavefunction of this state is given by a textbook formula:
\beq
\tran x{\psi_k} = \lp 2^{2k} (k!)^2 \pi\xi^2 \rp ^{-1/4}
                  e^{-x^2/2\xi^2} H_k(x/\xi), 
\label{hpp}
\eeq
where $H_k$ is the {\it k\/}th Hermite polynomial, and
\beq
\xi^2 = \lp {k_1 \over k_1 + k_2} \rp^{1/2} \equiv \tanh (\kap/2).
\label{xi}
\eeq
It is also useful to write
$\xi = e^{-\tta}$, in terms of which we have the symmetrical relations
\beq
\begin{array}{rcl}
e^{-2\tta} &=& \tanh(\kap/2), \\
e^{-\kap}  &=& \tanh\tta.
\end{array}
\label{tta}
\eeq

The number states $\ket n$ on the other hand, have wavefunctions
\beq
\tran xn = \lp 2^{2n} (n!)^2 \pi \rp ^{-1/4}
                  e^{-x^2/2} H_n(x). 
\label{hpn}
\eeq
We can thus evaluate the overlap $\tran n{\psi_k} = \bnk$ in the $x$ representation. 
The requisite integrals are most simply found by using the generating function
for the Hermite polynomials,
\beq
e^{2xz -z^2} = \sum_{n=0}^{\infty} {z^n\over n!} H_n(x).
\label{genf}
\eeq

The final result for $\bnk$ is
\beq
\bnk = \cos(\pi D) \lp {n!\,k! \over 2^{n+k} \cosh\tta} \rp^{1/2}
        \,\sum_{j=0}^{[k/2]}\,
         {(-1)^j \,(\tanh\tta)^{D+2j} \, (2\,\sech\tta)^{k-2j}
            \over
           (D+j)!\, (k-2j)! \, j!}. 
\label{bnkbig}
\eeq
Here $D\equiv (n-k)/2$, and $[x]$ denotes the largest integer less than or equal
to $x$. Note that the $\cos(\pi D)$ factor accounts for the vanishing of $\bnk$
if $n-k$ is odd. The latter fact is a simple consequence
of the parity relation $H_n(-x) = (-1)^n H_n(x)$. It can
be seen as a restatement of the separation of $\ham$ into two subspaces,  or of
the $\Dta m =2$ step in Eq.~(\ref{Seq}), and rigorously implies $\cmk = 0$
for odd $m + S - k$ for the entire range of $m$.

Next, let us find $\cmk$ in the central or classically forbidden region
where $S-m$ and $S+m$ are both $O(S)$. In this region, we have to sufficient
accuracy
\bea
w(m) &=& \left( \kc \rp
         \left( \blp\sph\brp^2 - m^2 \rp + O(S^0), \\
   \label{wm2}
t(m) &=& {\tst{1 \over 4}} k_2\left( \blp\sph\brp^2 - m^2 \rp
                + O(S^0). \label{tm2}
\eea
It is easy to see that $d(w/t)/dm = O(S^{-1})$, so the quasiclassicality
condition (\ref{slowq}) holds if $S \gg 1$. Equation (\ref{qj}) then yields
for the local wavevector
\bea
\cos q(m) &=& {2\e0k - (2k_1 + k_2) \left( \blp\sph\brp^2 - m^2 \rp
                 \over
               k_2 \left( \blp\sph\brp^2 - m^2 \rp} \nnu \\
          &=& -\left( {2k_1 + k_2 \over  k_2} \rp +
              {2\e0k \over k_2(2S+1)}
              \left( {1\over \smm} + {1\over \spm } \rp.
\label{qm2}
\eea
For the discrete WKB method to work, the energy levels which are split must lie
well below the barrier, i.e., $\e0k \ll (k_1 + k_2/2)S^2$. The second term in
Eq.~(\ref{qm2}) is then of order $S^{-1}$ relative to the first, and we can
solve for $q(m)$ as an expansion in powers of $1/S$. Equation (\ref{xi}) implies that
\beq
1 + 2k_1/k_2 = \cosh \kap, \label{kap} 
\eeq
so that
\beq
q(m) = \pi + i \left[ \kap - {\e0k\over \om_0}
               \left( {1\over S +{1\over2} -m} + {1\over S +{1\over2} + m} \rp
              + O \left( {\e0k \over \om_0 S^2} \rp \right].
\label{qmasy}
\eeq
The exponent in Eq.~(\ref{Cj}) therefore equals (recalling the extra factor of
$1/2$ due to the step $\Dta m=2$)
\beq
\phi(m) = i\int^m q(m') {dm'\over 2}
        = {i\pi m\over 2} - {\kap m\over 2} +
             {\e0k\over 2\om_0}\ln 
               \left( {S+{1\over2} +m}\over {S+{1\over2}-m} \rp + \cdots.
\label{phim}
\eeq

Similarly, correct to leading order in $S$, we have
\beq
|v(m)| = {\om_0\over 2S} \left( \blp\sph\brp^2 - m^2 \rp
\label{vm}
\eeq

We now substitute Eqs.~(\ref{phim}) and (\ref{vm}) into Eq.~(\ref{Cj}), changing
the normalization from unit flux to total unit probability in accord with the
discussion in Sec.~III. Writing the result only for the nonzero $\cmk$, i.e., for
$m+S-k$ even, we have
\beq
\cmk \approx A_k {(S+{1\over2} +m)^{(2k-1)/4} 
                         \over (S+{1\over2} -m)^{(2k+3)/4}}
                e^{i\pi(m-k)/2}\,e^{-\kap m/2},
\label{cmid}
\eeq
where $A_k$ is the normalization constant, and we have used $\e0k/\om_0 = k+\hf$.

It is now possible to match Eqs.~(\ref{cmid}) and (\ref{bnkbig}) directly,
since both results are valid in the region $S \gg n \gg 1$. 
We first find $\bnk$ for $S \gg n \gg k$, with $k=O(1)$:
\beq
\bnk = {\cos (\pi(n-k)/2) \over \sqrt{k!}}
       \left( {2\over\pi} \rp^{1/4}
            (\sech\tta)^{(2k + 1)/2}
           (\tanh\tta)^{(n-k)/2} n^{(2k - 1)/4}.
\label{bedge}
\eeq
To find $\cmk$ in the overlap region, we put $m=-S + n$ in Eq.~(\ref{cmid}), 
recall that $e^{-\kap} = \tanh\tta$, and in this way obtain for $n\gg1$,
\beq
C^{(k)}_{-S+n} = {A_k \over (2S)^{(2k+3)/4}}
                 \cos(\pi(n-k)/2) e^{-i\pi S/2}
                  (\tanh\tta)^{(n-S)/2} n^{(2k -1)/4}.
\label{Cint}
\eeq
Comparing with Eq.~(\ref{bedge}) we obtain
\beq
A_k = e^{i\pi S/2} \left( {2\over\pi} \rp^{1/4}
      {1\over \sqrt{k!}}
           (\sech\tta)^{(2k + 1)/2}
           (\tanh\tta)^{(S-k)/2} (2S)^{(2k +3)/4}.
\label{Ak}
\eeq

The last step is to apply Eq.~(\ref{edif}) for the splitting. The only
issue requiring caution is that of even versus odd $N$, the number of
points on which the wavefunction $\cmk$ is nonzero. We have $N = S+1$
for even $k$ and $N=S$ for odd $k$. Thus the parity of $N$ is that of $S+k-1$.
Both cases can be easily handled at the same time,
however. From Eq.~(\ref{cmid}), we
see that for $|m| \ll S$, the nonvanishing $\cmk$'s are given by
$ (A_k/S)e^{-\kap m/2}$ up to an overall phase.
Thus, the factor multiplying $2\,\tmid$ in Eq.~(\ref{edif}) equals
$2\sinh\kap\,|A_k/S|^2$ in magnitude in both cases. Similarly, we
see from Eq.~(\ref{tm}) that $\tmid$ equals $k_2 S^2/4$ in both cases. Thus,
\beq
\Dk = k_2 \sinh\kap\,|A_k|^2
\label{Dk}
\eeq
for all $k$ and $S$.

It remains to substitute Eq.~(\ref{Ak}) into Eq.~(\ref{Dk}). We quote the
answer in terms of the parameters $\om_0$ and $\tta$:
\bea
\Dk &=& {1\over k!} \lp {4S \over \sinh 2\tta} \rp^k \D0, \label{Dkf} \\
\D0 &=& 4\,\om_0 \lp {S\over\pi} \rp^{1/2} \sech\tta (\tanh\tta)^S. \label{D0f}
\eea
For some purposes, it may be desirable to write these results in terms of the
ratio $\xi$ defined in Eq.~(\ref{xi}). The conversion is easily done using the
formulas
\bea
\tanh\tta &=& (1-\xi^2)/(1+\xi^2), \label{ttta} \\
\sech\tta &=& 2\xi/(1+\xi^2),      \label{stta} \\
(1/\!\sinh 2\tta) &=& 2\xi/(1-\xi^4). \label{s2tta}
\eea

As mentioned before, the result (\ref{D0f}) for the ground state splitting
agrees with that of
Enz and Schilling~\cite{es2} and Belinicher, Providencia, and Providencia~\cite{bpp}.
A comparison with numerically obtained exact answers is given in Table I. As is
evident, the agreement is quite good, and improves with increasing $S$.

The formula (\ref{Dkf}) for the higher splitings is, to our knowledge, new.

\section{Tunnel splitting for model Hamiltonian (1.5)}

In this section we apply our method to the Hamiltonian (1.5), limiting ourselves
to finding the ground state tunnel splitting $\D0$. This time the recursion
relation connects contiguous coefficients $C_{m-1}$, $C_m$, and $C_{m+1}$, and
\bea
w(m) &=&  -\gam m^2, \label{w2m} \\
t(m) &=&  (-\al /2)(S'^2 - m^2)^{1/2}, \label{t2m}
\eea
where we have introduced the convenient abbreviation
\beq
S' = S + 1/2. \label{Sp}
\eeq

Let us begin, as in the previous section, by finding the wavefunction for the
state localized in the right hand well, $C_m$, in the classically
accessible region. This time the answer depends on how the parameters $\al$
and $\gam$ are taken to vary as $S \to \infty$. To see this note that the lower
band-edge function $U^-(m)$ now has 
a quadratic as opposed to linear form near its minima (see Fig. 2).
If we define
\beq
m = S' \cos\tta, \label{msp}
\eeq
then
\beq
U^-(m) = -\gam S'^2 \cos^2\tta - \al S' \sin\tta. \label{Ut}
\eeq
The mimima are located at $\tta = \tta_1$, $\pi - \tta_1$, where
\beq
\sin\tta_1 = \al/2\gam S'. \label{tta1}
\eeq
It follows that if $\al$ and $\gam$ remain fixed as $S \to \infty$, the
minima approach the points $m = \pm S$, which gives us a problem similar to
that of the previous section. For novelty's sake, let us therefore take $\al$
and $\gam$ to scale so that $\tta_1$ remains fixed and not close to 0 as
$S \to \infty$. We shall see that our final answer for $\D0$ will hold even when
this condition is not satisfied. 

For future use,
let us define $\tta_0$ as the angle at which the classical expression for
the energy (not $U^-(m)$) has a minimum. We have
\beq
\sin\tta_0 = \al/2\gam S. \label{tta0}
\eeq
Further,
\beq
\sin\tta_1 = \lp 1 - (1/2S) + \cdots \rp \sin\tta_0; \quad
  \cos\tta_1 = \cos\tta_0 + \sin^2\tta_0/2S\cos\tta_0 + \cdots.
   \label{tta10}
\eeq

With the above conditions on $\tta_1$,
it is easy to find $C_m$ near the potential minima.
Let us define
\beq
m_1 = S' \cos\tta_1. \label{m1}
\eeq
The conditions on $\al$ and $\gam$ imply that $S-m_1 = O(S)$. Since the wavefunction
we seek is essentially
that for the ground state of a harmonic oscillator, its spatial extent,
$\Dta m$, on the other hand, is of order $S^{1/2}$. This can be seen by expanding the
semiclassical Hamiltonian in powers of $q$ and $m-m_1$. We have
\bea
\ham_{\rm sc}(q,m) & = & w(m) + 2 t(m) \cos q \nnu \\ 
                   & \approx & -\gam S'^2(1+\sin^2\tta_1) +
                     \gam S'^2 \sin^2\tta_1 q^2 + 
                      \gam\cot^2\tta_1 (m-m_1)^2 + \cdots. \label{hquad}
\eea
The leading neglected terms are, ignoring $S'$ independent constants, of order
$S'(m-m_1) q^2 \equiv h_1$, $S'^2 q^4 \equiv h_2$, and $(m-m_1)^3/S' \equiv h_3$.
It follows that the small oscillation frequency $\om_0$ and the ground state energy
$E_0$ (in the absence of tunneling, needless to say) are given by
\bea
\om_0 &=& 2\gam S' \cos\tta_1, \label{om2} \\
E_0 &=& -\gam S'^2 (1 + \sin^2\tta_1) + \gam S' \cos\tta_1. \label{eg2}
\eea
The wavefunction can be written down directly as for a harmonic oscillator with
a continuous spatial coordinate,
\beq
C_m = \left( {\cos\tta_1 \over \pi S' \sin^2\tta_1} \rp^{1/4}
      \exp \left( - {\cos\tta_1 \over 2S' \sin^2\tta_1} (m-m_1)^2 \rp,
\label{wfcl}
\eeq
which is already normalized to unity inside the well.

What is the range of validity of Eq.~(\ref{wfcl})? This can be found by demanding that
the terms ignored, $h_1$, $h_2$, and $h_3$, be much smaller than $\om_0/2$, the value of
the $q^2$ and $(m-m_1)^2$ terms in Eq.~(\ref{hquad}). Since the kinetic and potential
energies are equal for a harmonic oscillator, we have $q^2 \propto (m-m_1)^2/S'^2$
in the region of validity. Thus, $h_1 = O(m-m_1)^3/S' = O(h_3)$, and
$h_2 = O(m-m_1)^4/S'^2 \ll h_1, h_3$. Demanding that $\om_0/2 \gg h_1$, we find that
Eq.~(\ref{wfcl}) is valid provided
\beq
|m-m_1| \ll (S')^{2/3}. \label{val}
\eeq
Note that this range keeps us clear of the end point $m=S$.

The second step is to find $C_m$ in the classically forbidden region, especially
near the point $m=0$ (if $S$ is integral) or $m=1/2$ (if $S$ is half-integral). 
Since $q(m)$ is imaginary, we define
\beq
q(m) = i\kap(m), \label{kap2m}
\eeq
with $\kap(m) > 0$. Using the basic definition (\ref{qj}) with the energy (\ref{eg2}),
we obtain
\beq
\cosh \kap(m) = {\sin^2\tta + \sin^2 \tta_1 \over 2\sin\tta \sin\tta_1}
                - {\cos\tta_1 \over 2 S' \sin\tta \sin\tta_1}
                + O(S')^{-2}.
\label{chkap}
\eeq
Writing
\beq
\kap(m) = \kap_0(m) + \kap_1(m) + \cdots, 
\label{kapex}
\eeq
where $\kap_j$ is of order $1/S'^j$, we obtain
\bea
\kap_0(m) &=& {1\over 2} \log \left( {S'^2 - m^2 \over S'^2\sin^2\tta_1} \rp,
    \label{k02} \\
\kap_1(m) &=& {1\over 2}\left[ {1\over m - m_1} - {1\over m + m_1} \right].
    \label{k12}
\eea
The exponent in Eq.~(\ref{Cj}) therefore equals
\bea
\phi(m) &=&  \phi_0(m) + \phi_1(m) \nnu \\
        &=&  {1\over 2} \left[ (S' + m) \log(S' + m) - (S' - m) \log(S'-m)
                              -2m\log(eS'\sin\tta_1) \right] \nnu \\
        &\quad&    +{1\over 2} \left[ \log(m_1 -m) - \log(m_1 + m) \right].
            \label{phi2m}
\eea
Here, $\phi_0$ and $\phi_1$ are the quantities given on the first and
second lines above, and correspond, resepctively, to $\kap_0$ and $\kap_1$.

The velocity is now easy to find. To leading order in $S$, we get
\beq
|v(m)| = 2\gam S'^2 \sin\tta_1 \sin\tta \times \sinh\kap_0 = \gam(m_1^2 - m^2).
\eeq
The wavefunction in the central region therefore equals
\beq
C_m = {A \over m + m_1} \exp\phi_0(m), \label{wfmid}
\eeq
where $A$ is a constant that must be found by matching the solution with
Eq.~(\ref{wfcl}). This in turn is done by expanding $\phi_0(m)$ in powers of
$m-m_1$, which is permissible as $|m-m_1| \ll S^{2/3}$ in the matching region.
We find
\bea
\phi_0(m) &=& \phi_{00} -{\cos\tta_1 \over 2 S' \sin^2\tta_1} (m-m_1)^2;
                \label{phi0} \\
\phi_{00} &=& S'[\log\cot(\tta_1/2) - \cos\tta_1]. \label{phi00}
\eea
Putting $m=m_1$ in the pre-exponential factor in Eq.~(\ref{wfmid}), we see that
the forbidden region $C_m$ has exactly the structure required to match on to
Eq.~(\ref{wfcl}). Comparing the two expressions, we obtain
\beq
A = 2 m_1 \lp {\cos\tta_1 \over \pi S' \sin^2\tta_1} \rp^{1/4} e^{-\phi_{00}}.
   \label{A2}
\eeq

The final step is to apply the formula (\ref{edif}) for the splitting. In doing this,
the following simplifications are useful. First,
\beq
\phi_0(m) = -m \log\sin\tta_1 \qquad {\rm for} \quad m = O(1). \label{phimid}
\eeq
Second, to leading order in $S$,
\beq
\om_0 = 2\gam S \cos\tta_0. \label{om02}
\eeq
Collecting together all these facts, we obtain, for both
integral and half-integral $S$,
\beq
\D0 = 4 \om_0 \lp {S\over \pi}\rp^{1/2} {\cos^{3/2}\tta_0 \over \sin\tta_0}
      \lp {1-\cos\tta_0 \over 1+ \cos\tta_0} \rp^{S + 1/2} e^{2S\cos\tta_0}.
\label{split2}
\eeq
This is the main result of this section. It can be seen to be identical to
Eq.~(16) of Enz and Schilling \cite{es2}, via the substitutions $A = \gam$,
$a = \sin\tta_0$ and $b=0+$. It is almost, but not quite identical to Eq.~(4.11)
of Garg and Kim \cite{gk}, which errs by a factor of $O(S^0)$.

Several comments about Eq.~(\ref{split2}) are in order. First, although we have
only derived it assuming that $\sin\tta_0 \gg 1/S$, it holds in fact for all
$\sin\tta_0$, provided of course that the WKB approximation itself is valid.
This can be seen from the numerical comparison we give below, and also analytically,
by assuming that $\sin\tta_0 = O(1/S)$ from the very start. This suggests that
Eq.~(\ref{split2}) holds uniformly for all $\sin\tta_0$ not too close to 1, but
we have not investigated this point carefully.
As a subcomment, note that in the limit $\tta_0 \to 0$, i.e., vanishing $\al$,
$\D0 \sim \al^{2S}$. This is of course, exactly what is expected from perturbation
theory --- it takes $2S$ applications of the perturbation $-\al S_x$ to connect
the states with $S_z = \pm S$ \cite{hb}.
Second, the WKB validity restriction means that the field $\al$ in Eq.~(\ref{ham2})
can not be too large. Quantitatively, we require that the energy barrier be
much greater than $\om_0$, the level spacing in the well. This condition can be
shown to be equivalent to
\beq
1 - \sin\tta_0 \gg S^{-2/3}. \label{bigbar}
\eeq 
Third, as long as Eq.~(\ref{bigbar}) holds, in the limit where the field is large
and the barrier is small, Eq.~(\ref{split2}) reduces to
\beq
\D0 = 4{\sqrt 3} \om_0 (B_0/2\pi)^{1/2} e^{-B_0}, \label{splquar}
\eeq
where $B_0 = (2/3) S \cos^3\tta_0 = (16/3)\Dta U/\om_0$, $\Dta U$ being the
energy barrier. This form is the well known answer for the tunnel spliting of
a massive particle moving in a quartic potential double well. The spin problem
can be transformed into such a problem by noting that (i) in this limit $S_x \approx S$,
so that the commutator $[S_y, S_z] = iS_x$ can be approximately reproduced by
writing $S_z =x$, $S_y = -Sp$, where $p$ and $x$ are momentum and position operators
with the commutator $[p,x] = -i$, and (ii)  expanding the square root in the relation
$S_x = \left[ S^2 - (S_y^2 + S_z^2)\right]^{1/2}$, and rewriting $S_y$ and $S_z$ in
terms of $p$ and $x$ as indicated above. We leave it to the reader to check that the
frequency $\om_0$ and energy barrier $\Dta U$ of this particle problem are given by
$2\gam S\cos\tta_0$ and $\gam S^2\cos^3\tta_0/4$, so that the answer for $\D0$
is indeed given by Eq.~(\ref{splquar}).

Fourth, let us compare our result for $\D0$ with the numerical and analytic
answers obtained by van Hemmen and S\"ut\H o \cite{vs2}, which apply when
$\sin\tta_0 = O(1/S)$. This comparison is shown in Table II. The parameters are
$\gam = \al =1$, and the two approximate forms for the splitting are given by
\beq
\Da = 2\gam S\exp[-2S \log(\gam S)], \label{dasy}
\eeq
\beq
\Di = {\om_0 \over 2}
        \exp\Biggl[ - \int_{-m_t}^{m_t} \cosh^{-1}
            \lp {-E_0 - \gam m^2 \over
                               \al \left[S(S+1) - m^2\right]^{1/2}} \rp \Biggr].
                   \label{dint}
\eeq
In Eq.~(\ref{dint}), $\pm m_t$ are the classical turning points, where $E_0 = w(m)$,
and $E_0$ is the numerically obtained ground state energy. van Hemmen and S\"ut\H o
describe $\Da$ as an  ``asymptotic" result, and $\Di$ as
the ``exact WKB" answer. As can be seen from the columns containing the ratios
of the splittings given by these formulas to the numerically found splittings, neither
result asymptotes to the correct answer. [Recall that two functions $f(x)$ and
$g(x)$ are said to be asymptotically equal as $x \to x_0$, i.e., $f(x) \sim g(x)$, if
$f(x)/g(x) \to 1$ as $x \to x_0$.] $\Da$
actually gets worse with increasing $S$, which indicates that
a multiplicative term such as $e^{-cS}$ has been omitted ($c$ is some constant),
while $\Di$ is off by a term of $S^0$ \cite{rt2}.
Neither result therefore meets
the goal set in Sec.~I of our paper: to find all terms $B_n$ in Eq.~(\ref{asy}) that
are of order $S^0$ or greater.
Equation (\ref{split2}), on the other hand, does just that. It is also obvious that
the recipe given by van Hemmen and S\"ut\H o [see Eq.~(5.19) of Ref. \onlinecite{vs2}]
cannot really be exact in general as regards the prefactor.

Lastly, Eq.~(\ref{split2}) can be compared with the answer obtained by Scharf,
Wreszinski, and van Hemmen \cite{SWv}, which is valid when $\sin\tta_0 = O(1/S)$.
Their result --- see their Eqs.~(1.7) and (1.8), and Table 1 --- is also off by
$O(S^0)$. In fact the discrepancy is again $(e/\pi)^{1/2}$, which is to be expected
in light of their method \cite{rt2}.

\acknowledgments
This work is supported by the NSF via grant number DMR-9616749.

\begin{figure}
\caption{Discrete WKB potential energy diagram for tunnel splitting problem. For
a particle with energy $E_0$, the region between $j_1$ and $j_2$ is classically
forbidden.}
\end{figure}
\begin{figure}
\caption{Sketch of the band-edge functions for the Hamiltonian (\ref{ham2}). The
key point is that $U^-(m)$ now has a quadratic minimum.}
\end{figure}

\begin{table}[t]
\caption{Comparison between numerical and analytical [Eq.~(\ref{D0f})]
results for the ground state tunnel splitting for the Hamiltonian (\ref{ham}).
Numbers in parentheses give the power of 10 multiplying the answer.
We have chosen $k_1 = 5.0/S$, $k_2 = 20.0/S$. This scaling keeps $\om_0$ fixed
as $S \to \infty$.}
\vspace{0.2cm}
\begin{center}
\begin{tabular}{c c c c}
$S$ & $\D0$ (numerical) & $\D0$ (analytic) & Error(\%) \\
\hline
10 & $9.282(-3)$ & $9.749(-3)$ & 5.0 \\
11 & $3.738(-3)$ & $3.906(-3)$ & 4.5 \\
12 & $1.497(-3)$ & $1.558(-3)$ & 4.1 \\
13 & $5.974(-4)$ & $6.195(-4)$ & 3.7 \\
14 & $2.375(-4)$ & $2.455(-4)$ & 3.4 \\
15 & $9.412(-5)$ & $9.708(-5)$ & 3.1 \\
16 & $3.721(-5)$ & $3.830(-5)$ & 2.9 \\
17 & $1.468(-5)$ & $1.508(-5)$ & 2.7 \\
18 & $5.778(-6)$ & $5.927(-6)$ & 2.6 \\
19 & $2.271(-6)$ & $2.326(-6)$ & 2.4 \\
20 & $8.910(-7)$ & $9.115(-7)$ & 2.3
\end{tabular}
\end{center}
\end{table}

\begin{table}[t]
\caption{Comparison between numerical and various analytical and
semianalytical results for the ground state tunnel splitting of the
Hamiltonian (\ref{ham2}). The parameters are $\al = \gam = 1$. As
in Table I, $x(n)$ denotes the number $x \times 10^n$. The
quantities $\Dta_n$, $\Da$, and $\Di$ are taken from Table I
of van Hemmen and \Suto  [19], and are, respectively, the numerically
computed splitting, the result (\ref{dasy}), and the numerically
evaluated integral expression (\ref{dint}). The quantity $\D0$ is
given by Eq.~(\ref{split2}) of this work.}
\vspace{0.2cm}
\begin{center}
\begin{tabular}{c c c c c c c c}
{}& Numerics & \multicolumn{4}{c}{Analytic/semianalytic results from Ref. \cite{vs2}} & 
             \multicolumn{2}{c}{This work} \\
$S$ & $\Dta_n$ & $\Da$\tablenote{The numbers have been given to greater accuracy
and corrected in a few places.}
 & $\Da/\Dta_n$ & $\Di$ & $\Di/\Dta_n$ & $\D0$ & $\D0/\Dta_n$ \\
\hline
3 & 1.44(-3) & 4.12(-3) & 2.86 & 2.01(-3) & 1.40 & 1.48(-3) & 1.03 \\
4 & 1.18(-5) & 6.10(-5) & 5.17 & 1.64(-5) & 1.39 & 1.20(-5) & 1.02 \\
5 & 5.18(-8) & 5.12(-7) & 9.88 & 7.25(-8) & 1.40 & 5.24(-8) & 1.01 \\
6 & 1.43(-10) & 2.76(-9) & 19.3 & 2.00(-10) & 1.40 & 1.44(-10) & 1.01 \\
7 & 2.68(-13) & 1.03(-11) & 38.5 & 3.77(-13) & 1.41 & 2.70(-13) & 1.01 \\
8 & 3.66(-16) & 2.84(-14) & 77.7 & 5.15(-16) & 1.41 & 3.68(-16) & 1.01 \\
9 & 3.79(-19) & 6.00(-17) & 158 & 5.36(-19) & 1.41 & 3.81(-19) & 1.01 \\
10 & 3.09(-22) & 1.00(-19) & 324 & 4.37(-22) & 1.41 & 3.10(-22) & 1.00 \\
11 & 2.03(-25) & 1.35(-22) & 666 & 2.87(-25) & 1.41 & 2.03(-25) & 1.00 
\end{tabular}
\end{center}
\end{table}


\begin{references}
\bibitem{rt2} Readers may wonder if all that is wrong with van Hemmen and S\"ut\H o's
formula is the typographical omission of a factor of $2^{-1/2}$. This is not so.
An inspection of the arguments accompanying their Eqs.~(5.19) and (C.10-12) reveals
that there is no scope for such a factor and it is purely accidental that the error
seems to be $2^{1/2}$ in this case. Indeed, use of a similar formula for potential
problems with massive particles in one dimension also yields incorrect prefactors;
for the quartic double well, e.g., the error is known to be $(e/\pi)^{1/2}$. The
same error is present in Scharf, Wreszinski, and van Hemmen's result \cite{SWv}.
\end{references}
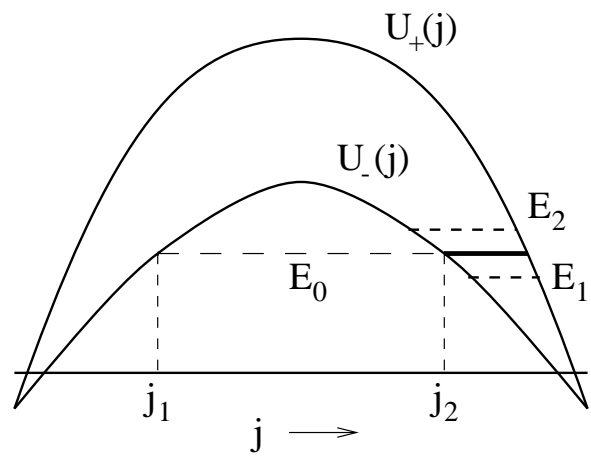
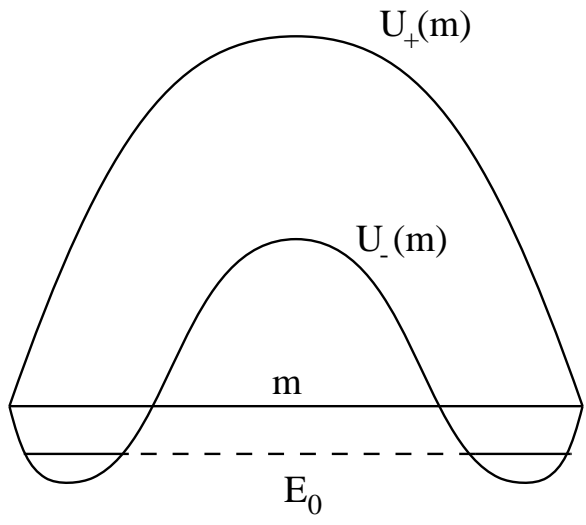
\end{document}